\documentclass[longauth]{aa}  

\usepackage{natbib,graphicx,amssymb,amsmath}
\bibpunct{(}{)}{;}{a}{}{,} 
\newcommand{\kms}{$\rm km\,s^{-1}$} 
 
\newcommand{\tfm}{\tablefootmark}
\newcommand{\tft}{\tablefoottext}

\begin{document}
\title{Multi-molecule ALMA observations towards the Seyfert 1 galaxy NGC~1097}
\titlerunning{Multi-molecule observations towards NGC~1097}
\author{
  S. Mart\'in \inst{\ref{inst1}} 
  \and K. Kohno \inst{\ref{inst2},\ref{inst3}} 
  \and T. Izumi \inst{\ref{inst2}} 
  \and M. Krips \inst{\ref{inst1}} 
  \and D. S. Meier \inst{\ref{inst4},\ref{inst5}} 
  \and R. Aladro \inst{\ref{inst12}} 
  \and S. Matsushita \inst{\ref{inst13}} 
  \and S. Takano \inst{\ref{inst6},\ref{inst7}} 
  \and J. L. Turner \inst{\ref{inst16}} 
  \and D. Espada \inst{\ref{inst8},\ref{inst9},\ref{inst10}}
  \and T. Nakajima \inst{\ref{inst11}} 
  \and Y. Terashima \inst{\ref{inst23}} 
  \and K. Fathi \inst{\ref{inst17},\ref{inst18}} 
  \and P.-Y. Hsieh \inst{\ref{inst13},\ref{inst14}} 
  \and M. Imanishi \inst{\ref{inst10},\ref{inst15}} 
  \and A. Lundgren \inst{\ref{inst8}} 
  \and N. Nakai \inst{\ref{inst22}} 
  \and E. Schinnerer \inst{\ref{inst20}} 
  \and K. Sheth \inst{\ref{inst19}} 
  \and T. Wiklind \inst{\ref{inst8}} 
}
\institute{
  Institut de Radio Astronomie Millim\'etrique, 300 rue de la Piscine, Dom. Univ., 38406, St. Martin d'H\`eres, France\\
  \email{smartin@iram.fr}\label{inst1}
  \and
  Institute of Astronomy, The University of Tokyo, 2-21-1 Osawa, Mitaka, Tokyo 181-0015, Japan\label{inst2}
  \and
  Research Center for the Early Universe, The University of Tokyo, 7-3-1 Hongo, Bunkyo, Tokyo 113-0033, Japan\label{inst3}
  \and
  New Mexico Institute of Mining and Technology, 801 Leroy Place, Socorro, NM, USA\label{inst4}
  \and 
  National Radio Astronomy Observatory, Pete V. Domenici Array Science Center, P.O. Box O, Socorro, NM, 87801, USA\label{inst5}
  \and
  European Southern Observatory, Alonso de C\'ordova 3107, Vitacura, Casilla 19001, Santiago 19, Chile\label{inst12}
  \and
  Academia Sinica, Institute of Astronomy \& Astrophysics, P.O. Box 23-141, Taipei 10617, Taiwan\label{inst13}
  \and
  Nobeyama Radio Observatory, Nobeyama, Minamimaki, Minamisaku, Nagano 384-1305, Japan\label{inst6}
  \and
  Department of Astronomical Science, The Graduate University for Advanced Studies (Sokendai), Nobeyama, Minamimaki, Minamisaku, Nagano 384-1305, Japan\label{inst7}
  \and
  Department of Physics and Astronomy, UCLA, 430 Portola Plaza, Los Angeles, CA 90095-1547, USA\label{inst16}
  \and
  Joint ALMA Observatory (JAO), Alonso de C\'ordova 3107, Vitacura, Santiago, Chile\label{inst8}
  \and
  National Astronomical Observatory of Japan (NAOJ), 2-21-1 Osawa, Mitaka, 181-8588, Tokyo, Japan\label{inst9}
  \and
  Department of Astronomical Science, The Graduate University for Advanced Studies (SOKENDAI), 2-21-1 Osawa, Mitaka, 181-8588, Tokyo, Japan\label{inst10}
  \and  
  The Solar-Terrestrial Environment Laboratory, Nagoya University, Furo-cho, Chikusa-ku, Nagoya, Aichi 464-8601, Japan\label{inst11}
  \and
  Department of Physics, Ehime University, 2-5 Bunkyo-cho, Matsuyama, Ehime 790-8577, Japan\label{inst23}
  \and
  Stockholm Observatory, Department of Astronomy, Stockholm University, AlbaNova Centre, 106 91 Stockholm, Sweden\label{inst17}
  \and
  Oskar Klein Centre for Cosmoparticle Physics, Stockholm University, 106 91 Stockholm, Sweden\label{inst18}
  \and
  Institute of Astronomy, National Central University, No. 300, Jhongda Road, Jhongli City, Taoyuan County 32001, Taiwan\label{inst14}
  \and
  Subaru Telescope, National Astronomical Observatory of Japan, 650 North A’ohoku Place, Hilo, HI 96720, USA\label{inst15}
  \and
  Division of Physics, Faculty of Pure and Applied Science, University of Tsukuba, Tsukuba, Ibaraki 305-8571, Japan\label{inst22}
  \and
  Max Planck Institute for Astronomy, K\"onigstuhl 17, Heidelberg 69117, Germany\label{inst20}
  \and
  National Radio Astronomy Observatory, 520 Edgemont Road, Charlottesville, VA 22903, USA\label{inst19}
}

\abstract
{
The nearby Sy~1 galaxy NGC~1097 represents an ideal laboratory to explore the molecular chemistry in the presence and surroundings of an active galactic nucleus (AGN).
}
{
Exploring the distribution of different molecular species allows us to understand the physical processes affecting the ISM both in the AGN vicinity as well as in the outer
star forming molecular ring.
}
{
We carried out 3~mm ALMA observations which include seven different molecular species, namely HCN, HCO$^+$, CCH, CS, HNCO, SiO, HC$_3$N, and SO as well as the $^{13}$C isotopologues of the first two.
Spectra were extracted from selected positions and all species were imaged over the central 2~kpc ($\sim 30''$) of the galaxy at a resolution of $\sim2.2''\times1.5''$ ($150~pc\times 100~pc$).
}
{
HCO$^+$ and CS appear to be slightly enhanced in the star forming ring. CCH, showing the largest variations across NGC~1097, is suggested to be a good tracer of both obscured and
early stage star formation. HNCO, SiO and HC$_3$N are significantly enhanced 
in the inner circumnuclear disk (CND) surrounding the AGN.
}
{
{Differences in the molecular abundances are observed between the star forming ring and the inner circumnuclear disk. 
We conclude that the HCN/HCO$^+$ and HCN/CS differences observed between AGN dominated and starburst (SB) galaxies are not due to a HCN enhancement due to X-rays, but rather this enhancement is produced by shocked
material at distances of $200$~pc from the AGN. Additionally we claim the lower HCN/CS to be a combination of a small under-abundance of CS in AGNs together with excitation
effects, where a high dense gas component ($\sim 10^6~\rm cm^{-3}$) may be more prominent in SB galaxies.
However the most promising are the differences found among the 
dense gas tracers which, at our modest spatial resolution, seem to outline the physical structure of the molecular disk around the AGN. In this picture, HNCO probes the well shielded
gas in the disk, surrounding the dense material moderately exposed to X-ray radiation traced by HC$_3$N. Finally SiO might be the innermost molecule in the disk structure.
}
}

\maketitle

\section{Introduction}
The molecular material in galaxies is a fundamental ingredient in the fueling of the energetic phenomena in galactic nuclei.
Though the mass and distribution of the molecular gas can be probed by the bright lines of carbon monoxide, deep observations of fainter molecular species
have been proven as excellent tools of the physical processes affecting the molecular gas.
This has been evidenced by numerous deep spectral line surveys during the last decade \citep[see review by][]{Mart'in2011a} as well as extensive work on the theoretical modeling
\citep[i.e.][]{Meijerink2007,Bayet2008a,Bayet2011,Loenen2008,Meijerink2013,Harada2010,Harada2013}.
The molecular chemistry in the ISM provides direct information on the environmental conditions and therefore on the type of activity in the central 
few hundred parsecs of galaxies.
Moreover, the observation of individual molecular species allow us not only to estimate the physical conditions, but to dissect the ISM into the
different heating mechanisms in play as observed within the central molecular zone of our own Galaxy \citep{Amo-Baladr'on2011,Jones2012,Mart'in2012}

Most of the deep and/or unbiased chemical studies in galaxies have been carried out at very low resolution \citep[i.e.$>>100$~pc,][]{Wang2004,Mart'in2006,Mart'in2011,Aladro2011a,Aladro2013,Davis2013,Watanabe2014}
but for absorption studies towards bright continuum high-z sources where, though extremely prolific in molecular detection,
no spatial information can be obtained \citep{Muller2011,Muller2014}.
High resolution studies towards extragalactic sources have mostly focused towards the brightest nearby galaxies as well as the brightest species after carbon monoxide, that is HCN and HCO$^+$
(i.e. 
NGC~253 at $3''$ resolution, \citet{Knudsen2007};
M51 at $4''$, \citet{Schinnerer2010}; 
NGC~1068 at $1''$, \citet{Krips2011}; 
NGC~1097 at $3''$, \citet{Hsieh2012};
or a sample of LIRGs at $2''-10''$, \citet{Imanishi2007,Imanishi2009}).
Imaging line surveys through interferometric
observations pushed our view of the ISM even further and has shown how the chemistry can provide unique insights on the variations of the physical processes across
the central kpc of starburst galaxies \citep{Meier2005,Meier2012}. However, such studies towards AGN galaxies have been more limited due mostly to the lack 
of sensitivity \citep[i.e.][]{Garc'ia-Burillo2010}.

The full potential of extragalactic chemistry is just emerging thanks to new generation instruments like ALMA and the upcoming NOEMA.
In fact, within the last year and while still in the early science phase, ALMA is already providing extremely high quality data on the chemistry of some 
of the brightest nearby prototypical galaxies, both AGNs (NGC~1068, \citet{Garcia-Burillo2014,Takano2014,Viti2014}; NGC~1097, \citet{Izumi2013}) and
SBs (NGC~253, Meier et. al submitted).
The unprecedented capabilities of these instruments allow us to extend molecular studies to either further or fainter objects, thus increasing the significance of chemical comparative studies, now
limited to a very reduced number of sources (Aladro et al submitted).

The prototypical nucleus in NGC~1097 hosts the first reported low-luminosity \citep[$L_{2-10keV}=4.4\times10^{40}~\rm ergs~s^{-1}$][]{Terashima2002,Nemmen2006}
Seyfert 1 nucleus \citep{Storchi-Bergmann1993}.
The nuclear region is surrounded by a $\sim700$~pc radius ring \citep{Barth1995,Quillen1995} with ongoing star formation at a rate of $\sim5M_\odot~\rm yr^{-1}$
\citep{Hummel1987}.
The molecular material in NGC~1097 has been extensively studied at high resolution \citep{Kohno2003,Hsieh2008,Hsieh2011,Hsieh2012}.
Molecular gas is distributed in a central concentration of $\sim 350$~pc that we will refer to as the circumnuclear disk, directly surrounding the
AGN, and the weaker 1.4~kpc diameter molecular ring , the circumnuclear star forming (SF) ring \citep{Hsieh2008}.

Though located at a similar distance \citep[14.5~Mpc,][]{Tully1988} than the Seyfert 2 galaxy NGC~1068, the nucleus of NGC~1097 does not only show a fainter X-ray luminosity, but it is also
one order of magnitude fainter both in IR luminosity \citep{Sanders2003} and dense gas as traced by HCN \citep{Hsieh2012,Krips2011}.
The unprecedented sensitivity of ALMA allows us to carry out multi-molecule imaging studies on NGC~1097 which can be directly
compared to similar studies towards the much brighter NGC~1068 \citep{Takano2014}.

In our first paper, ALMA observations of HCN and HCO$^+$ at both $J=4-3$ and $1-0$ were reported by \citet{Izumi2013}. 
In this paper we present a multi-molecule imaging study that makes use of the species that were simultaneously observed with the $J=1-0$ of HCN and HCO$^+$.
A companion paper will present the continuum emission in the 3mm band as well as a kinematic study of the ratios of the brightest species, while here we will only
discuss the overall integrated molecular emission.
Our aim is to provide insights on the variations of the molecular abundances in the $300$~pc region surrounding its Seyfert 1 nucleus,
as compared to the star forming dominated circumnuclear ring at a distance of $600-800$~pc from the center.

\section{Observations}
\begin{figure*}
\centering
\includegraphics[width=\linewidth]{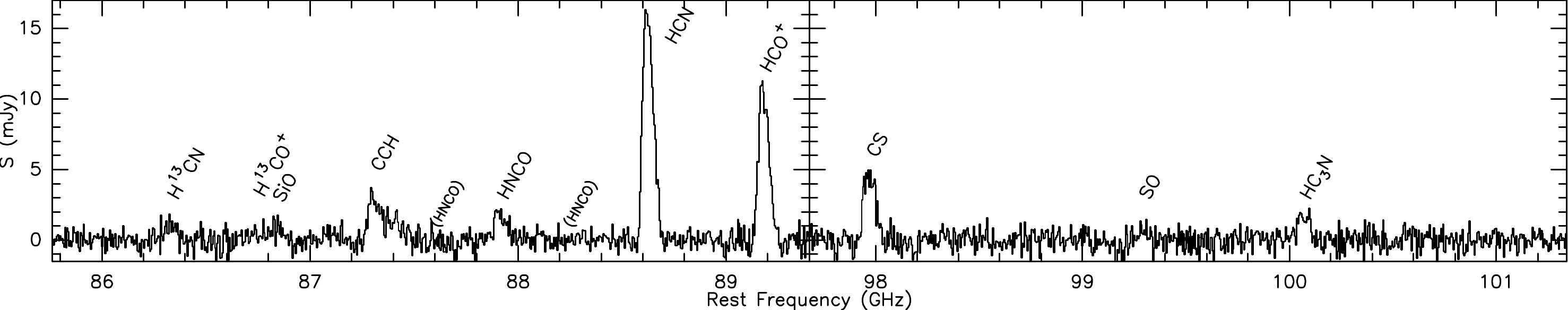}
\caption{Full spectral coverage of the ALMA Band 3 observations extracted from the nuclear position of NGC~1097 where the peak of molecular emission is observed for all species. 
This position corresponds to position A in Table~\ref{tab.selpos} and Fig.~\ref{fig.allmolecules}.
 \label{fig.pos1spec}}
\end{figure*}

3~mm wavelength observations towards NGC~1097 were carried out as part of the ALMA Cycle 0 early science program 2011.0.00108.S
(P.I. K. Kohno). Data were acquired during 2012 July 29th and October 19th with an overall on source integration time of $\sim 72$~min.
The dual sideband (2SB) Band 3 receivers were tuned at an LO1 frequency of 93.187~GHz. The correlator was configured in dual polarization frequency division mode (FDM)
with the $4\times1.875$~GHz spectral windows covering the rest frequency ranges $85.8-89.4$~GHz and $97.7-101.3$~GHz (Fig.~\ref{fig.pos1spec}).
The original frequency resolution of $488$~kHz was degraded by 10 channels down to a velocity resolution of $\sim17-14.5$~\kms~ across the covered frequency range.

The phase center of the observations was $\alpha_{J2000}=02^h46^m19.06^s$, $\delta_{J2000}=-30^\circ16'29.7''$.
The HPBW of the primary beam of the 12~m antennas is $72''-61''$ at the extremes of the frequency coverage.
At the distance of 14.5~Mpc, the angular to linear scale is $\sim70~pc/''$, therefore our observations covered 
the central $5-4.3$~kpc at half power.
The 24 and 31 antennas used on the first and second run, respectively, were configured up to distances of $\sim250$~m from the array center, thus covering
uv ranges from 4.3 to 132~$k\lambda$ ($14-447$~m).
This configuration resulted on an average synthesized beam of $\sim2.2''\times1.5''$ ($150~pc\times 100~pc$) with $P.A.=100^\circ$.

The quasars J1924-292 and J0334-401 were used as bandpass and phase calibrator, respectively. Absolute flux scaled was calibrated through observations towards
Neptune.
Data were calibrated and cleaned with CASA \citep{McMullin2007}, while images were produced with GILDAS \footnote{http://www.iram.fr/IRAMFR/GILDAS}.
A final rms of $\sim0.5$~mJy in the smoothed 4.88~MHz channels (i.e. 10 original channels) was measured in the final images.

\section{Results}
\label{sec.results}
As shown by the spectrum in Fig.~\ref{fig.pos1spec}, extracted from the peak of molecular emission position (see below), the $\sim8$~GHz frequency coverage of our observations allowed
the detection of 9 molecular species. For the identified features, Table~\ref{tab.specparams} presents some basic spectral parameters extracted from JPL spectral
line catalog \citep{Pickett1998}. 
Two non-detected transitions of HNCO (shown in brackets) are included as their limits will serve to set constraints to the excitation conditions (see Section~\ref{subsec.CD}).

\begin{table}
\caption{Spectral parameters of detected transitions \label{tab.specparams}}
\begin{tabular}{l c c l l}
\hline\hline
Molecule                 & Transition          &  Frequency         &   $E_{upper}$ \\
                         &                     &    GHz             &      (K)      \\
\hline
$\rm H^{13}CN$           &  $1-0$              &  86.340~\tfm{a}    &       4       \\
$\rm H^{13}CO^+$         &  $1-0$              &  86.754            &       4       \\
SiO                      &  $2-1$              &  86.846            &       6       \\
CCH                      &  $1-0$              &  87.316~\tfm{a}    &       4       \\
(HNCO)                   &  $4_{1,4}-3_{1,3}$  & 87.597             &      54       \\
HNCO                     &  $4_{0,4}-3_{0,3}$  & 87.925~\tfm{a}     &      10       \\
(HNCO)                   &  $4_{1,3}-3_{1,2}$  & 88.239             &      54       \\
HCN                      &  $1-0$              &  88.631~\tfm{a}    &       4       \\
$\rm HCO^+$              &  $1-0$              &  89.188            &       4       \\
CS                       &  $2-1$              &  97.980            &       7       \\
SO                       &  $2_3-1_2$          &  99.300            &       9       \\
HC$_3$N                  &  $11-10$            & 100.076            &      29       \\                        
\hline                                      
\end{tabular}                               
\tablefoot{Undetected species are shown in brackets.
\tft{a}{Line multiplicity is taken into account during the spectral fit based on JPL catalog entries. We refer to the frequency of the brightest component.}
}
\end{table}

\begin{figure*}
\centering
\includegraphics[width=\linewidth]{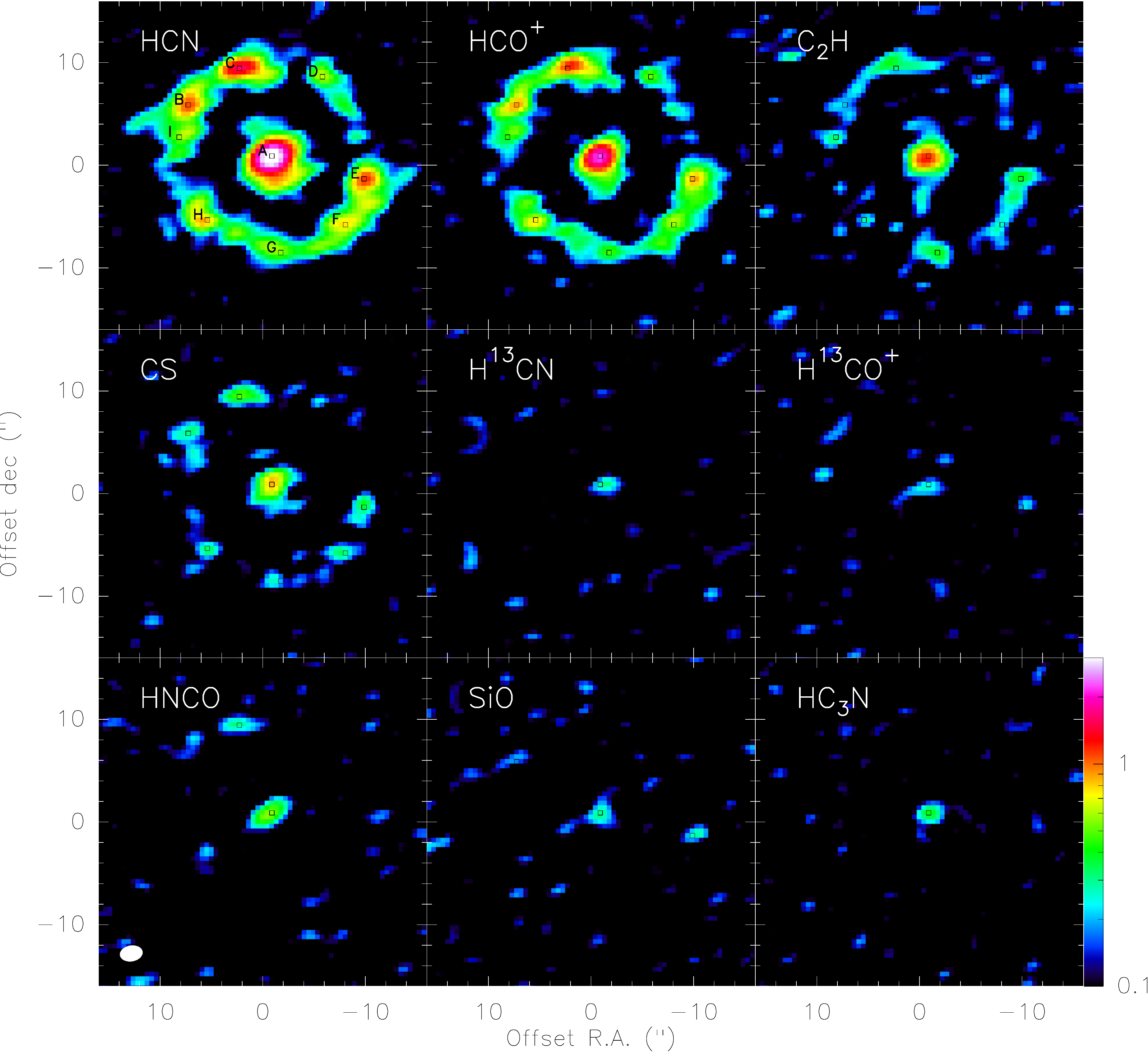}
\caption{Moment 0 maps of all detected species in the velocity range [$900-1600$]~\kms, except for CCH where a range [$600-1600$]~\kms~ was used to include the two groups
of hyperfine components. The $1~\sigma$ noise level in the integrated maps is 62~mJy~\kms~ for all species, but for CCH which is 77~mJy~\kms.
Logarithmic color scale is set to allow a direct comparison of the emission of the different species. Colour levels are set to a minimum of 0.1~mJy~\kms~ so that emission
over $3~\sigma$ level is seen above the black background. Positions where spectra were extracted are marked in all boxes, and labeled in the upper left box.
 \label{fig.allmolecules}}
\end{figure*}

Fig.~\ref{fig.allmolecules} presents the moment 0 maps of each of the detected species except for SO which does not stand out in the integrated maps (see discussion in
Sect.~\ref{subsec.CD}).
Moment 0 maps were integrated in the velocity range [$900-1600$]~\kms, where the molecular emission across the whole map is detected.
This range was used for all species but CCH, for which the velocity range [$600-1600$]~\kms~ (referred to the brightest hyperfine component) was
integrated in order to include the two groups of hyperfine transitions.
The common logarithmic color scale in Fig.~\ref{fig.allmolecules} allows a direct comparison of the emission of the different molecular species.

\subsection{Selected positions in AGN and SF ring}
To study the molecular abundance variations both in the circumnuclear disk around the AGN and the star forming ring, 
we have selected a sample of 9 positions.
Table~\ref{tab.selpos} lists the relative coordinates of the selected positions referred to the phase center of the observations.

Position A is selected as the brightest HCN emission position in the nuclear region of the NGC~1097 and it is offset from the phase center by $(-0.9'',0.9'')$, measured at the
center of the peak flux pixel in the image (with $0.45''$/pixel).
This position matches the measured position of the nucleus at 6~cm \citep[$\alpha_{J2000}=02^h46^m18.96^s$,$\delta_{J2000}=-30^\circ16'28.9'$;][]{Hummel1987} as well
as the $860~\mu m$ continuum peak \citep{Izumi2013}.
The spectra extracted at this position is shown in Fig.~\ref{fig.pos1spec}.

\begin{figure*}
\centering
\includegraphics[width=0.9\linewidth]{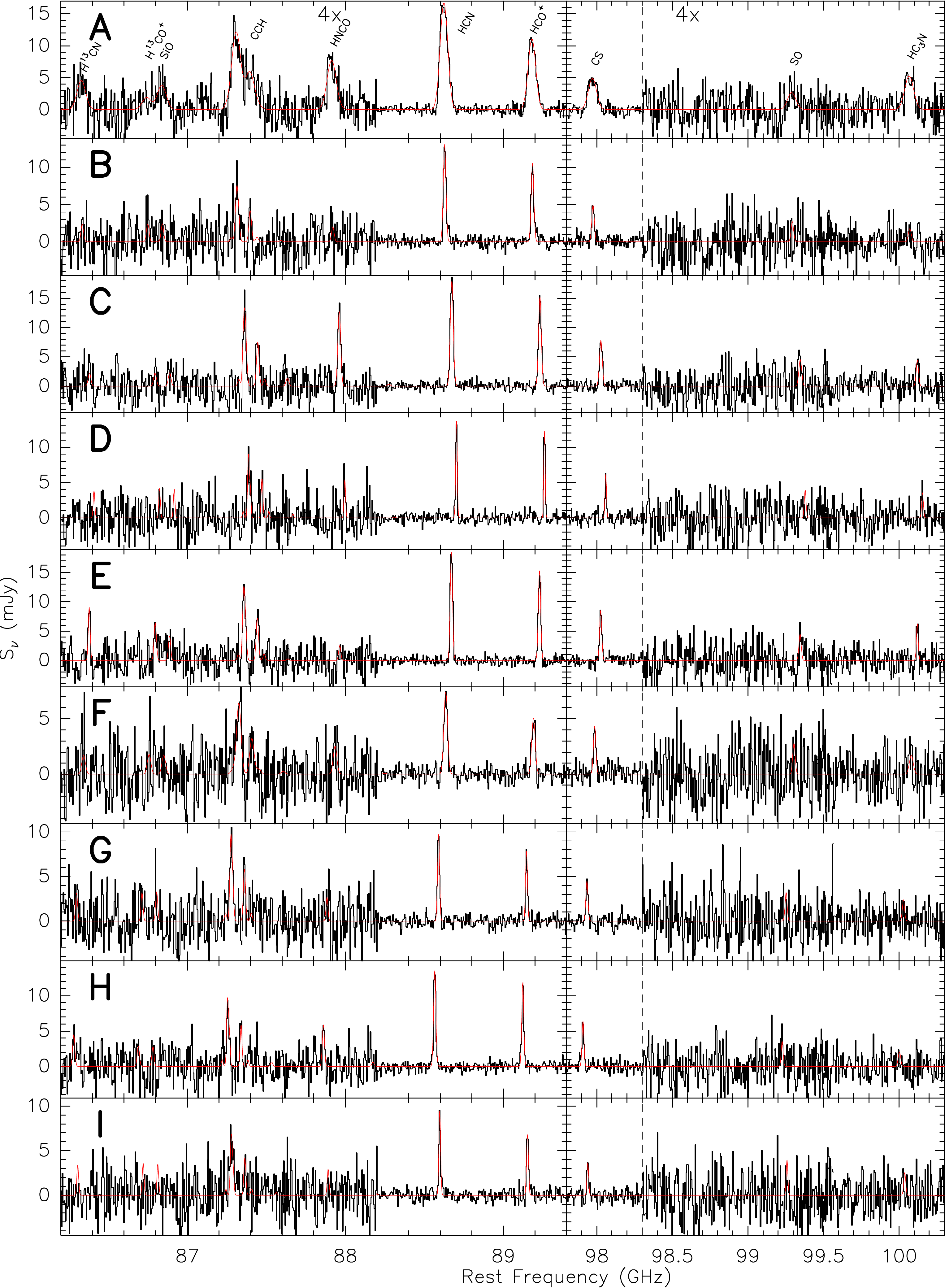}
\caption{Spectra extracted from selected positions in the AGN position (A) and on the SF ring (B-I) of NGC~1097. 
Frequency span has been cropped with respect to that of Fig~\ref{fig.pos1spec} removing the line free frequencies at the edges of both sidebands. 
Flux density scale has been multiplied by a factor of $\times 4$ in the regions (defined by the vertical dashed lines) containing the fainter lines to improve their visualization.
Molecular fit to the observed emission is represented by the red line. Non-detections are presented as the $3~\sigma$ limit Gaussian profiles.
 \label{fig.pos2to9spec}}
\end{figure*}

All other positions correspond to peaks of molecular emission of the brightest species, with the selection mainly driven by the maps of the brightest species,
namely HCN, C$_2$H and CS.
Fig.~\ref{fig.pos2to9spec} presents the individual spectra extracted from all selected positions, both in the AGN and in the SF ring. 

\begin{table}
\caption{Coordinates and line parameters of extracted spectra \label{tab.selpos}}
\begin{tabular}{l c c l l}
\hline\hline
Pos.       & $\Delta\alpha$& $\Delta\delta$ & $v_{\rm LSR}$     & $\Delta v_{1/2}$    \\
           &  ($''$)        &  ($''$)      &  (km\,s$^{-1}$)   & (km\,s$^{-1}$)      \\
\hline
A          &    -0.9        &   0.9        & 1290   (2)        & 190  (5)            \\
B          &     7.2        &   5.9        & 1276   (2)        & 65   (4)            \\
C          &     2.3        &   9.4        & 1222.2 (1.5)      & 74.5 (1.5)          \\
D          &    -5.8        &   8.6        & 1023.1 (1.0)      & 39   (2)            \\
E          &    -9.9        &  -1.3        & 1131.1 (0.6)      & 57.6 (1.4)          \\
F          &    -8.1        &  -5.8        & 1252   (3)        & 91   (6)            \\
G          &    -1.8        &  -8.5        & 1405.2 (1.2)      & 49   (3)            \\
H          &     5.4        &  -5.3        & 1485.9 (0.8)      & 56   (2)            \\
I          &     8.1        &   2.7        & 1380.3 (1.2)      & 43   (3)            \\
\hline    
\end{tabular}
\tablefoot{Relative coordinates refer to the observations nominal phase center $\alpha_{J2000}=02^h46^m19.06^s$, $\delta_{J2000}=-30^\circ16'29.7''$.
The errors derived from the fit for $v_{\rm LSR}$ and $\Delta v_{1/2}$ are indicated.}
\end{table}

\subsection{Excitation and column density determination}
\label{subsec.CD}
Rather than basing our result on direct integrated intensity line ratios, we have estimated the column densities of each of the molecular species
to have a more accurate determination of the molecular abundances in each position.

We have used the spectral line modeling tools within the MADCUBA\_IJ package\footnote{http://cab.inta-csic.es/madcuba/Portada.html} (Mart\'in et al in prep.)
to fit the molecular emission in the spectra of the selected sample positions.
The main assumption of this modeling is the emission under LTE conditions.
The parameters of column density, velocity and linewidth are left as free parameters.
Table~\ref{tab.selpos} presents the average velocity and linewidth measured in all species. See \citet{Hsieh2011} for a discussion on the kinematics and
the different linewidths measured in the star forming ring cloud complexes. 
However, excitation temperature and the source size parameters were fixed
to $8$~K and $1.8''$, respectively. These assumptions are explained in the following.

The gas in both the SF ring and around the AGN have kinetic temperatures higher than our assumed excitation temperature \citep[$T_k=30-90~\rm K$;][]{Pinol-Ferrer2011}.
This is due to the fact that the molecular gas is subthermally excited.
For HCN and HCO$^+$, \citet{Izumi2013} combined the $J=1-0$ and $4-3$ transitions, observed with ALMA, smoothed down to the resolution of the $3-2$ 
transition, observed with the SMA \citep{Hsieh2012}, in order to estimate the excitation conditions of the gas traced by these species. 
The emission can be roughly fitted in LTE by a rotational temperature of $\sim 8$~K.
This result is similar in the central position A around the AGN and towards the SF ring.
Though the assumption on the excitation temperature will not strongly affect the relative abundances given that most of our observed transitions have 
similarly low ($4-10$~K) energies, it might be an issue for the transition of HC$_3$N with a significantly higher upper energy level (Table~\ref{tab.specparams}).
However, the work by \citet{Aladro2011} towards nearby starbursts shows that such subthermal excitation affects equally to other molecular species in their 
lower $J$ transitions, not only to those at lower energies but up to transitions with$\sim 50$~K upper energy levels.
Additionally the non detection of the two higher energy transitions of HNCO listed in Table~\ref{tab.specparams} and shown in Fig.~\ref{fig.pos1spec}
set a limit on the excitation temperature of this species to $T_{ex}<20$~K.

Regarding the source size we assume the emission to be distributed in a region approximately equal to the synthesized beam in solid angle, 
that is $1.8''$ or $\sim125$~pc.
A smaller emitting region would directly scale up the column density, thus increasing the effect of opacity. 
Our fit to the spectra results in an opacity of $\tau\sim0.2$ in the brightest positions (A and B) for HCN.
Assuming a source size of half our assumed value would increase the opacities up to $\tau\gtrsim1$ and the derived column densities by a factor of $\sim 5$.
\citet{Izumi2013} also found low opacities in the high-J transitions of HCN in the center of NGC~1097.
Based on the peak line ratio of HCN and H$^{13}$CN of $\sim15$ and assuming a $^{12}$C/$^{13}$C isotopic ratio of $\sim 40$ \citep{Henkel2014} we can estimate
an opacity of $\tau>2$ towards the central position. It could be even larger for a higher carbon isotopic ratio \citep{Mart'in2010a}.
This shows that in the central position, the HCN emission either stems from regions significantly smaller than the beam, or the carbon isotopic ratio
is significantly smaller in this region.
We cannot disentangle these two possibilities with our data.
Though this uncertainty on the source size may have a moderate impact in the relative abundances of the brightest species, we have no further clues on the actual
molecular clump source sizes across the galaxy so we assume a similar source size for all positions as a first approximation.

The derived column densities and upper limits are listed in Table~\ref{tab.params} for all selected positions and detected species.
We note that both H$^{13}$CO$^+$ and SO detections are at the edge of detection. SO does not even show up in the integrated maps (Sect.~\ref{sec.results}), while the moment 0
detection of H$^{13}$CO$^+$ might be the result of SiO contamination.
Derived column densities for these two species are $\sim3\sigma$ in the few positions where those are detected, and it results in an overabundance
of H$^{13}$CO$^+$ in position E of a factor of $\sim4$ which may indicate the fit to be dominated by an spurious spectral artifact (see fit in Fig~\ref{fig.pos2to9spec}).
Thus these species will not be included in the following discussion. 

\begin{table*}
\caption{Column densities for each species and the selected position \label{tab.params}}
\centering
\scriptsize
\begin{tabular}{l l l l l l l l l l l}
\hline\hline
Pos.                & $N_{\rm HCN}$         & $N_{\rm HCO^+}$       & $N_{\rm CCH}$         & $N_{\rm CS}$          & $N_{\rm H^{13}CN}$    & $N_{\rm H^{13}CO^+}$  & $N_{\rm HNCO}$         & $N_{\rm SiO}$         & $N_{\rm HC_3N}$       & $N_{\rm SO}$       \\
                    & $\rm 10^{14}cm^{-2}$  & $\rm 10^{14}cm^{-2}$  & $\rm 10^{14}cm^{-2}$  & $\rm 10^{14}cm^{-2}$  & $\rm 10^{14}cm^{-2}$  & $\rm 10^{14}cm^{-2}$  & $\rm 10^{14}cm^{-2}$   & $\rm 10^{14}cm^{-2}$  & $\rm 10^{14}cm^{-2}$  & $\rm 10^{14}cm^{-2}$  \\
\hline
A                   & 3.14 (0.07)           & 1.16 (0.03)           & 14.4 (0.8)            & 1.85 (0.11)           & 0.26 (0.05)           & 0.06 (0.02)           &  3.1 (0.3)             & 0.25 (0.05)           &  2.6 (0.3)            & 0.6 (0.2)             \\
B                   & 0.84 (0.04)           & 0.38 (0.02)           & 3.3  (0.5)            & 0.56 (0.05)           & $<$0.04               & $<$0.02               & $<$0.2                 & $<$0.04               & $<$0.3                & $<$0.2                \\
C                   & 1.34 (0.02)           & 0.60 (0.02)           & 5.7  (0.6)            & 1.02 (0.06)           & $<$0.04               & $<$0.02               & 1.58 (0.17)            & $<$0.04               & 0.58 (0.15)           & 0.4 (0.1)             \\
D                   & 0.54 (0.03)           & 0.24 (0.01)           & 2.9  (0.4)            & 0.42 (0.03)           & $<$0.04               & $<$0.02               & 0.38 (0.09)            & $<$0.04               & 0.34 (0.11)           & $<$0.2                \\
E                   & 1.08 (0.02)           & 0.51 (0.02)           & 5.2  (0.5)            & 0.88 (0.04)           &  0.12 (0.02)          & 0.08 (0.02)           & $<$0.2                 & 0.07 (0.02)           & 0.62 (0.14)           & 0.3 (0.1)             \\
F                   & 0.64 (0.04)           & 0.24 (0.02)           & 4.0  (0.6)            & 0.48 (0.04)           & $<$0.04               & $<$0.02               &  0.4 (0.2)             & $<$0.04               & 0.37 (0.19)           & $<$0.2                \\
G                   & 0.47 (0.02)           & 0.20 (0.02)           & 3.7  (0.4)            & 0.43 (0.03)           & $<$0.04               & $<$0.02               & $<$0.2                 & $<$0.04               & $<$0.3                & $<$0.2                \\
H                   & 0.76 (0.02)           & 0.34 (0.02)           & 3.4  (0.4)            & 0.56 (0.06)           &  0.06 (0.02)          & $<$0.02               & 0.54 (0.14)            & $<$0.04               & $<$0.3                & $<$0.2                \\
I                   & 0.40 (0.02)           & 0.18 (0.02)           & 2.4  (0.4)            & 0.28 (0.05)           & $<$0.04               & $<$0.02               & $<$0.2                 & $<$0.04               & $<$0.3                & $<$0.2                \\
\hline
\end{tabular}
\end{table*}

\subsection{Relative molecular abundances}
We aim to understand the chemical differentiation between the molecular gas in the SF ring and the material
in the close vicinity of the nuclear engine.
The central few hundred parsecs CND contains a significant fraction of the dense molecular gas in NGC~1097 (Table~\ref{tab.params}).
Based on the integrated intensity maps of HCN and HCO$^+$ in Fig.~\ref{fig.allmolecules} the central 600~pc region encloses $32\%$ and $25\%$ of the total
detected emission, respectively.
This fraction is as much as $40\%$ for CS or CCH.
However, we are mostly interested in the relative abundances between the different molecular species so to determine whether an AGN driven chemistry can be
claimed in the central region.

\begin{figure*}
\centering
\includegraphics[width=\linewidth]{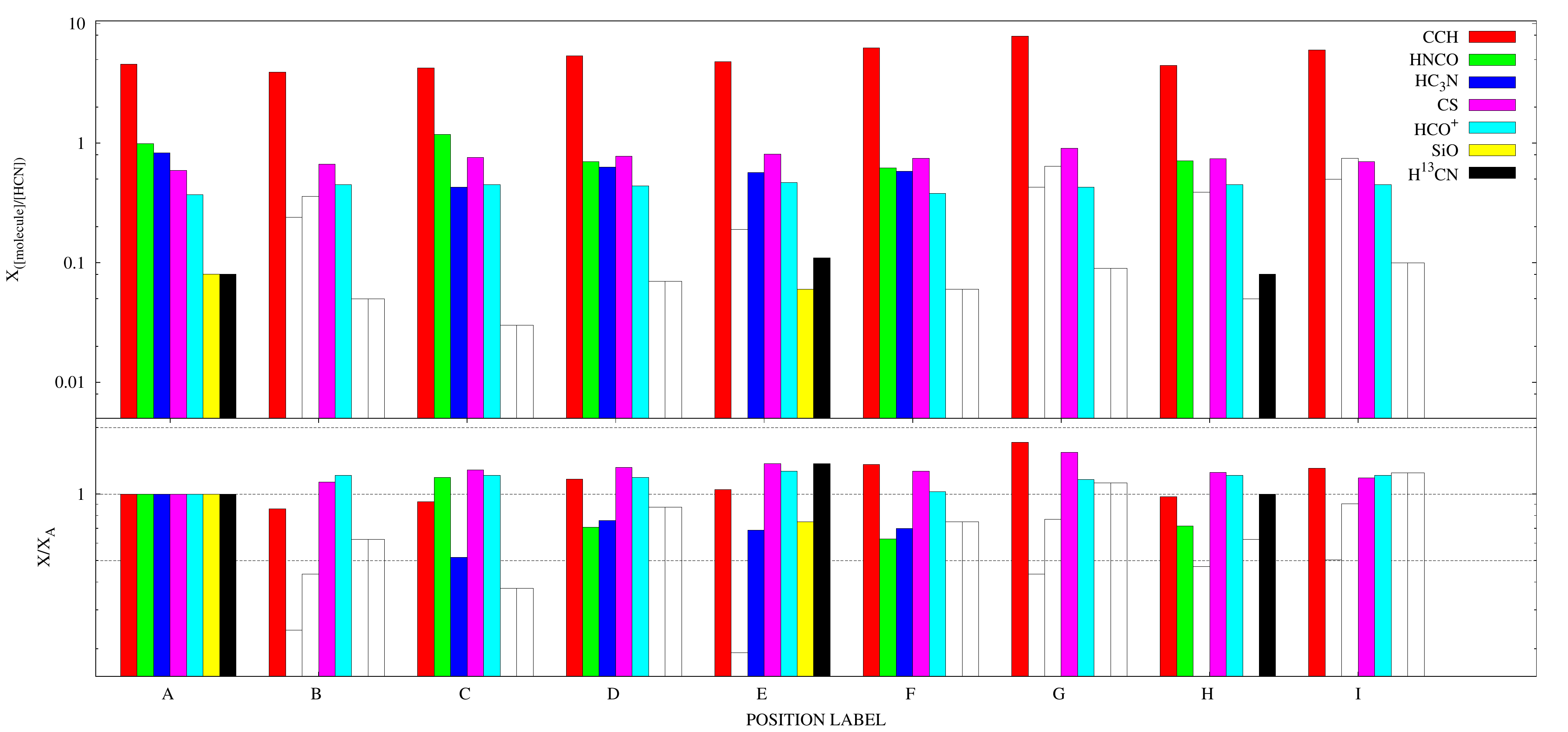}
\caption{(TOP) Molecular abundances relative to HCN estimated in the sample of selected positions for all detected species. Colored histograms represent species
detected while empty histograms represent upper limits to the detection. (BOTTOM) Abundances normalized
to those of position A for comparison. As a reference, the three horizontal dotted lines represent equal relative abundances or a factor of 2 above and below those in position A.
 \label{fig.RelativeHCN}}
\end{figure*}

We have used the brightest species in our observations, namely HCN, as a reference column density baseline. 
Fig.~\ref{fig.RelativeHCN} shows the abundances of each species relative to HCN. Molecules are displayed in decreasing order of abundance as measured in position A.
To better evaluate the differences between position A and those in the SF ring, we have normalized the relative abundances to those in position A (lower
panel in Fig.~\ref{fig.RelativeHCN}).

The selection of a good ``baseline'' species for normalization is important but not critical in our study.
As noted by \citet{Kohno2003}, HCN is observed to be significantly enhanced towards the nucleus relative to CO which might point out HCN as a not the ideal candidate to become
a reference molecule. However, this HCN/CO trend gets less evident when higher J transitions of CO are considered \citep{Hsieh2012}.

We see in Fig.~\ref{fig.RelativeHCN} that CS is consistently slightly more abundant in the star forming ring positions.
This could be caused by the normalization by HCN, implying an slight overabundance of HCN in the nuclear region as previously claimed as an AGN chemical imprint.
Therefore, another possibility among the brightest species would have been the use of CS, as done in previous works
\citep{Mart'in2009,Mart'in2011,Aladro2011a,Aladro2013}.
However, we note that reproducing Fig.~\ref{fig.RelativeHCN} with the abundances relative to CS results in an overall underabundances of all species detected in the SF ring compared to position A.
Given that, a priori, we have no reasons to assume all species to be overabundant in the CND, we stick to our use of HCN as it does provide abundances less biased towards the AGN position.
However, this uncertainty in the reference species is taken into account in the following discussion.

Though we will deal with relative abundance among species, for completeness we used the molecular hydrogen column densities derived from CO observations by \citet{Hsieh2011}
to estimate the fractional abundances relative to H$_2$.
On average the estimated fractional abundances of HCN would be $X(HCN)\gtrsim\times1-2\times10^{-9}$
both in the nuclear region and throughout the SF ring.
We consider this a limit given that they use the conversion factor $X_{CO}=3\times10^{20}\rm cm^{-2}~(K~km~s^{-1})^{-1}$ which is known to be overestimating of the H$_2$ column density
in dense environments. Thus the relative abundance could be as high as $X(HCN)\sim10^{-8}$.

\section{SF ring versus AGN abundances}

We find that the overall relative molecular abundances in the SF ring positions are in significantly good agreement with those in the nuclear region within less than a factor
of 2. The only exception is 
the underabundance of HNCO in some SF ring positions.
Though we do not find enormous differences in the abundances among the selected positions, some species appear to show a differentiation between their abundances in the surroundings of the AGN
and the positions in the star forming dominated ring.
In particular, HC$_3$N and HNCO appear to be the species showing the largest enhancement in the nuclear region.
In the following we further discuss 
the differences observed.

\subsection{HCO$^+$ and CS}
The relative abundances of HCO$^+$ and CS derived in the individual SF ring positions are found to be consistently, 
but still marginally regarding the excitation uncertainties,
$\sim20-50\%$ above those found towards the AGN position.

These two molecules are claimed to be enhanced by UV radiation in the surroundings of star forming regions.
Though its abundance may be equally enhanced by X-ray irradiation, HCO$^+$ is shown to be abundant in the presence of UV fields \citep{Bayet2008a,Bayet2011} and therefore it could
be associated to the massive star formation.
Additionally, HCO$^+$ is also observed to be enhanced in high velocity shocked material as observed in Galactic supernovae remnants \citep{Dickinson1980,Wootten1981} which, again
would link its enhancement to massive star formation.
High abundances of CS may be understood through the enhancement of its precursor S$^+$ in the gas phase\citep{Drdla1989,Goicoechea2006}, though there is some disagreement between theoretical
models and observations \citep{Mart'in2008,Tideswell2010}.

Both HCO$^+$ and CS follow closely the emission of HCN, and therefore they may be just probing the same dense gas.
We do not observe a clear morphological differentiation with respect to HCN at our achieved resolution in the SF ring so as to attribute their
emission to the star formation in the region.
However, the significant star formation in the ring  could explain their overall significantly enhanced emission with respect to the central region.
We observe an HCN/HCO$^+$ ratio which is higher, both in the AGN and the SF ring, than what is measured in other starburst galaxies \citep{Krips2008}.
Rather than a general enhancement of HCN in this NGC~1097 due to the presence of the AGN, it is likely that in starbursting galaxies HCO$^+$ is further enhanced due to the 
star formation event.
In such SF dominated environments, ionization due to cosmic rays accelerated in supernovae remnants \citep{Ackermann2013}
would also contribute to pull HCN/HCO$^+$ down \citep{Wootten1981} compared to what is observe even in the moderately SF ring on NGC~1097, or similarly in NGC~1068.
This scenario may as well explain why this ratio is moderately lower in the SF ring than in the very nuclear region, while the highest HCN/HCO$^+$ ratio would be a local effect in the central region
as discussed in the following. 

\subsubsection{AGN Irradiation?: HCN/HCO$^+$ ratio}
\label{HCN_HCO_SFAGN}
For more than a decade now, it has been claimed that the X-ray irradiation in AGN dominated galaxies results in an enhancement of the HCN molecule relative to the
overall molecular gas traced by CO or that affected by star formation traced by HCO$^+$ \citep{Kohno2001,Krips2008}.
In terms of relative abundances, we find that the $X(\rm HCN/HCO^+)=2.7\pm0.1$ in the central position is only marginally larger than the values in the SF ring which range
$2.1-2.7$.

\begin{figure}
\centering
\includegraphics[width=\linewidth]{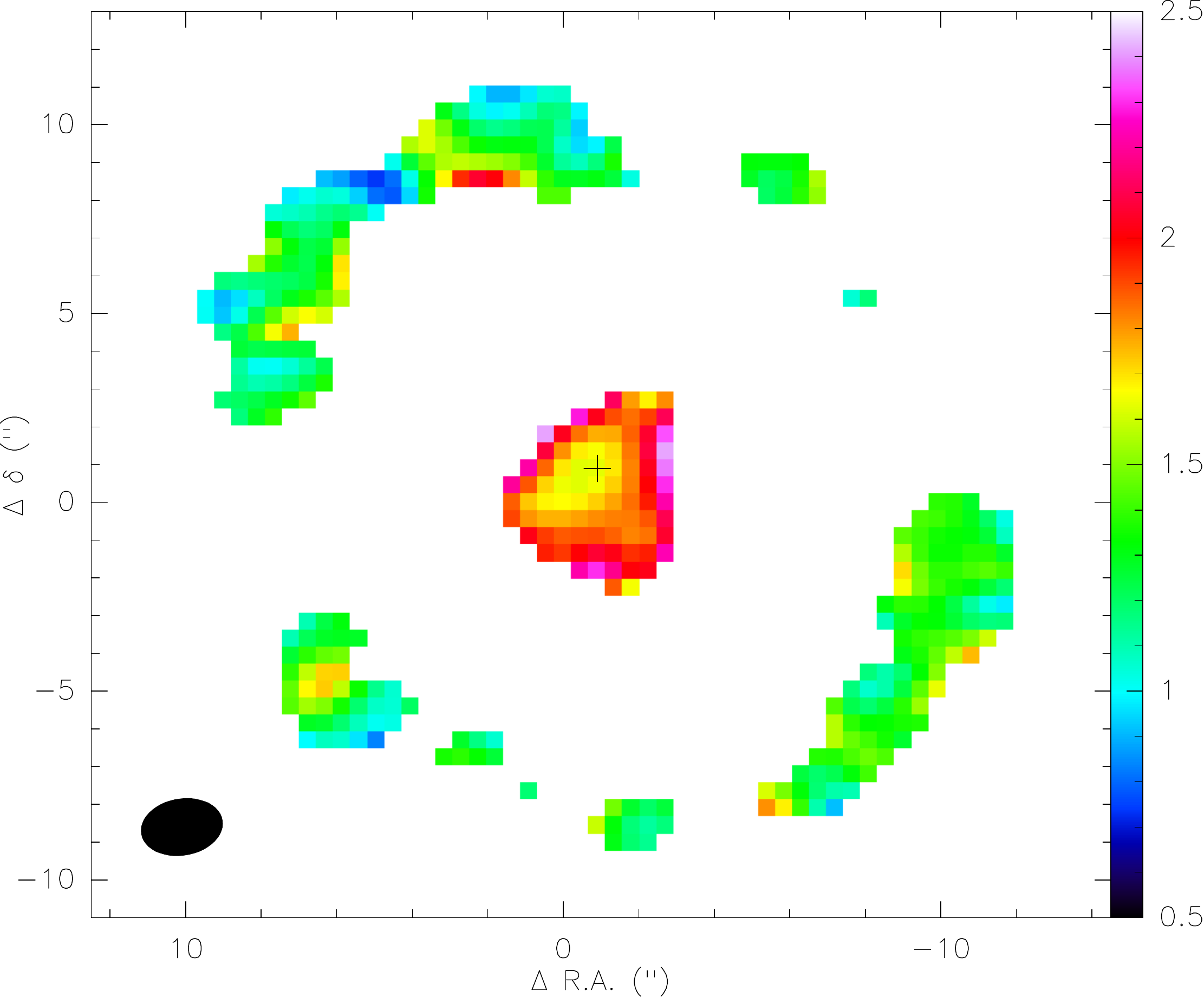}
\caption{
Integrated intensity HCN-to-HCO$^+$ line ratio. Values only above $5~\sigma$ in the individual integrated images has been used in this map.
The position of the peak of molecular emission is shown with a cross. Synthesized beam size is represented in the bottom left corner.
\label{fig.RatioHCNHCOp}}
\end{figure}

The high sensitivity of the ALMA data allows us to study the distribution of this ratio across the imaged region.
In Fig.~\ref{fig.RatioHCNHCOp} we show HCN/HCO$^+$ integrated line intensity ratio,
following discussions in the literature.
To make the most out of the high quality of the data and to avoid noise effects, we have only included points above a $5~\sigma$ level in the individual images.

Regarding its comparison with the outer SF ring, a clear HCN/HCO$^+$ integrated intensity ratio enhancement is observed in the central 300~pc region of NGC~1097,
as expected in an AGN dominated environment. However, this enhancement does not reach the maximum at the central position, as discussed in Sect.~\ref{Sect.AGNHCNHCOp}.
It is interesting to note that the HCN/HCO$^+$ measured is consistently high ($>1$) all over the galaxy. 
That is, setting these values in the proposed diagnostic diagrams based on the HCN/HCO$^+$ ratio \citep{Krips2008}, all the molecular gas in NGC~1097
would appear in the AGN dominated regime.
The difference between the HCN/HCO$^+$ in the nucleus as compared to the SF ring is less pronounced than in NGC~1068 where the ratio drops to $\sim1$ in the
ring \citep{Viti2014}. However, they only detected these species in a handful of compact positions on the SF ring, and a significant fraction of the gas in the image might
be resolved out.

Extensive literature can be found where HCN is claimed to be enhanced by the effect of X-ray irradiation 
\citep[see][and references therein]{Gracia-Carpio2006,Krips2008}. 
A clear correlation is found between the CO-normalized X-ray flux in the $6-8$~keV band and HCN emission \citep{Garcia-Burillo2014}, which is, however, also found for HCO$^+$,
SiO and CN \citep{Garc'ia-Burillo2010}.
Possibilities for the HCN enhancement such as oxygen depletion are discarded based on ancillary molecular
line ratios not matching the predictions \citep{Usero2004}.
The enhancement due to the presence of massive star forming regions explored by \citet{Gracia-Carpio2006} for LIRGs was discarded towards NGC~1097
in view of the low star forming rate in its nuclear region \citep{Izumi2013}.
However, even the X-ray enhancement appears to be out of the question \citep{Izumi2013} given that most models assume AGN luminosities of $L_X\sim10^{43-44}~\rm erg~s^{-1}$
\citep{Meijerink2005,Meijerink2007,Meijerink2013}, which is more than two orders of magnitude higher than the values measured in NGC~1097 or NGC~1068.
Nevertheless, intrinsic X-ray fluxes may be significantly higher. Such is the case of NGC~1068 where its nucleus is absorbed by
H$_2$ column density $>10^{24}\rm cm^{-2}$ \citep{Matt1997} and its intrinsic luminosity is estimated to be a $\sim2$ orders of magnitud larger than observed \citep{Iwasawa1997,Colbert2002}.
However, in the case of NGC~1097, the absence of Fe K emission lines in its X-ray spectra completely 
rules out the possibility of the presence of a heavily absorbed AGN revealing an intrinsically low-luminosity AGN.
In such case
it is not expected that the sphere of AGN influence goes beyond a few tens of parsecs in these sources.
Fig.~\ref{fig.RatioHCNHCOp} shows the central region of HCN/HCO$^+$ enhancement to be extended well beyond one synthesized beam (see Section~\ref{Sect.AGNHCNHCOp}).
Based on the modelling by \citet{Harada2010}, \cite{Izumi2013} therefore proposed high-temperature chemistry as the key to enhance HCN and therefore obtain an increase in the HCN/HCO$^+$ ratio in the nucleus
of NGC~1097.
LVG analysis by \citet{Hsieh2008} based on interferometric CO $2-1$ and $1-0$ shows the presence of such high temperature gas ($T_{kin}>400$~K) in the nuclear region.
Similarly, the LVG analysis of a sample of AGN sources with enhanced HCN emission resulted in kinetic temperatures $T_K>40$~K \citep{Krips2008} at single dish resolution.
High resolution multi-transition observations towards NGC~1068 show clumps in the CND with temperatures above 100~K \citep{Krips2011,Viti2014}.
Regarding the origin of such heating, mechanical heating is suggested to explain the required high temperature in NGC~1097 \citep{Loenen2008,Izumi2013}.

A more detailed exploration of the variation of this ratio as the gas approaches the central engine is found in Section~\ref{Sect.AGN}.

\subsubsection{The HCN/CS diagnostic}

\begin{figure}
\centering
\includegraphics[width=\linewidth]{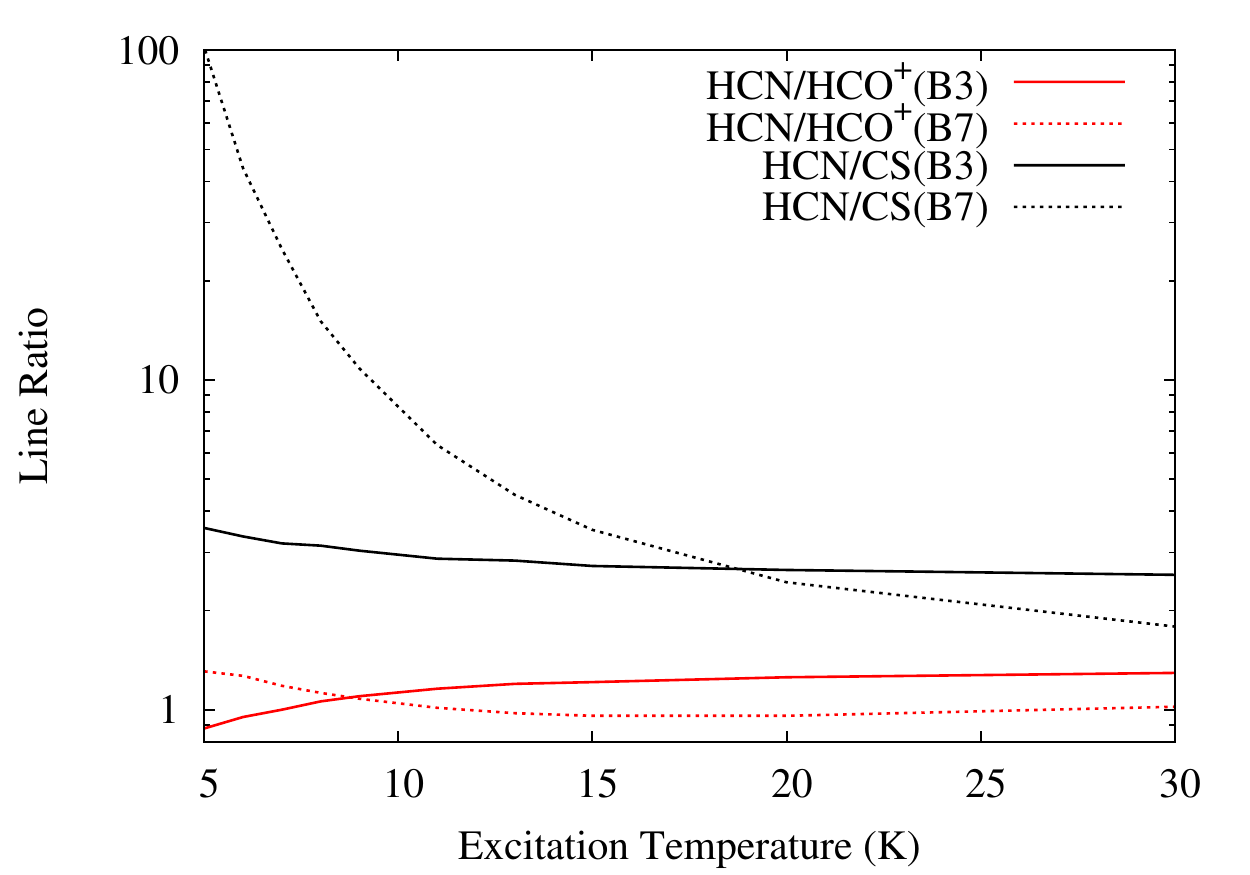}
\caption{
HCN/HCO$^+$ and HCN/CS ratios calculated for the observable transitions in the ALMA band 3 
(HCN and HCO$^+$ $1-0$, and CS $2-1$) and band 7 (HCN and HCO$^+$ $J=4-3$, and  CS $J=7-6$) as a function
of the LTE excitation temperature. The curves have been calculated to reproduce the observed integrated line ratios for $T_{\rm ex}=$8~K.
\label{fig.LineRatiosTex}}
\end{figure}

Recently, \citet{Izumi2013} proposed an AGN vs SB diagnostic diagram based on the HCN($J=4-3$)/CS($J=7-6$) ratio.
Towards NGC~1097, where the CS line was not detected down to a level of $2.6$~mJy, this ratio was reported to be a factor of $>3$ larger 
than in starburst galaxies.
This difference in the line ratio in AGN versus SB galaxies has also been reported at the lower-$J$ transitions with single dish observations towards NGC~1068 compared to those
in NGC~253 and M~82 \citep{Nakajima2011,Aladro2013}.
Based on our 5~mJy~beam$^{-1}$ detection of CS $J=2-1$ and assuming the LTE at $T_{\rm ex}=8$~K, the CS $J=7-6$ line would have a peak flux density of $\sim 0.6$~mJy~beam$^{-1}$.
The fact of measuring a significantly different HCN/CS ratio between the high-$J$ ($>13$) and the low-$J$ transitions ($\sim3$) is the result of the difference between the upper energy
levels of the transitions involved in the ratios.
While the HCN and CS ALMA band 3 transitions have similar upper energy levels (Table.~\ref{tab.specparams}), there is a significant difference between that of HCN $J=4-3$ (42~K) and
CS $J=7-6$ (65~K).
Such difference makes the HCN($J=4-3$)/CS($J=7-6$) ratio very sensitive to variations in the excitation temperature defining the energy levels populations.
This is illustrated in Fig.~\ref{fig.LineRatiosTex} where we calculated the HCN/HCO$^+$ and HCN/CS line ratios for the transitions observable in the ALMA band 3 and 7 bands.
While HCN/HCO$^+$ ratios are basically insensitive to excitation conditions (as long as both species share the same excitation), similar to the low-$J$ HCN/CS ratio,
the high-$J$ HCN/CS ratio shows a strong variation as a function of the excitation temperature.
Thus the ratio among the high-$J$ HCN/CS is a less reliable probe to the abundance variations if the excitation conditions are not well constraint.

Rather than rejecting the diagnostic proposed by \citet{Izumi2013} based on this ratio, our result aims at understanding the origin of the large ratio difference observed between AGN and SB
galaxies.
Though a small abundance difference of up to a factor of 2 may be claimed as is suggested by the observed difference in the low-$J$ transition ratio \citep{Aladro2013}, the larger
difference observed in the high-$J$ ratio is likely due to the excitation of CS.
In starburst galaxies, the high-$J$ transitions of CS appear to be associated to a moderately higher excitation temperature gas component \citep[$\sim15-30$~K,][]{Aladro2011}.
This component appears to correspond to a dense molecular component ($\sim10^6\rm cm^{-3}$).
From the results in NGC~1097, we can suggest that this dense molecular component may be less prominent in AGNs, or CS might be not so efficiently produced there.
However a more complete high resolution study of the excitation of the gas in NGC~1097 and NGC~1068 should help understanding the origin of this underabundance and/or underexcitation.

\subsection{Star formation: CCH}
\label{Sect.SF}
CCH is observed to vary significantly throughout the SF ring, with measured relative abundances $20\%$ below (position B), up to  a $70\%$ above (position G) those in the center.
The distribution of the emission and abundances of CCH differs significantly from other species in NGC~1097.

Within the Galaxy this species is observed to be located in the UV irradiated cloud edges associated with 
massive star forming regions \citep{Beuther2008,Walsh2010,Li2012}.
Its formation paths are favored in photodissociation regions (PDRs) by the reaction of C$^+$ with small hydrocarbons and additionally through photodissociation of C$_2$H$_2$ \citep[][and references therein]{Meier2005}.
However UV radiation is also its main destruction path, and therefore CCH abundance is dependent on the evolutionary stage of the star forming
event \citep{Li2012}.
In external galaxies the abundance of CCH appears not to be affected by the type of activity \citep{Nakajima2011,Aladro2013}.
However, observations towards an extended sample of galaxies shown that CCH in AGNs may be similartly overabundant to starburst in the late stages of evoluton
while being different to earlier stage starbursts (Aladro et al. submitted).
Overall, when the average galactic emission is observed in a sample of galaxies, a good correlation between CCH and HCO$^+$ is found \citep{Martin2014}.
However, we do not find such correlation at GMC complex scale in our data.
At high resolution, observations towards starburst galaxies \citep[][Meier et al. submitted]{Meier2005,Meier2012} have shown CCH to be tracing UV irradiated clouds in the 
vicinity of both nuclear stars clusters and embedded star forming regions, showing large abundance variations across these galaxies.
Similarly, we observe large relative abundances variations both in the nucleus and the SF ring in NGC~1097. Such variations might explain the observed homogeneity between
different activity types where, when averaged over the whole few hundred parsecs of a galaxy, the local variations are smoothed out.

\begin{figure}
\centering
\includegraphics[width=\linewidth]{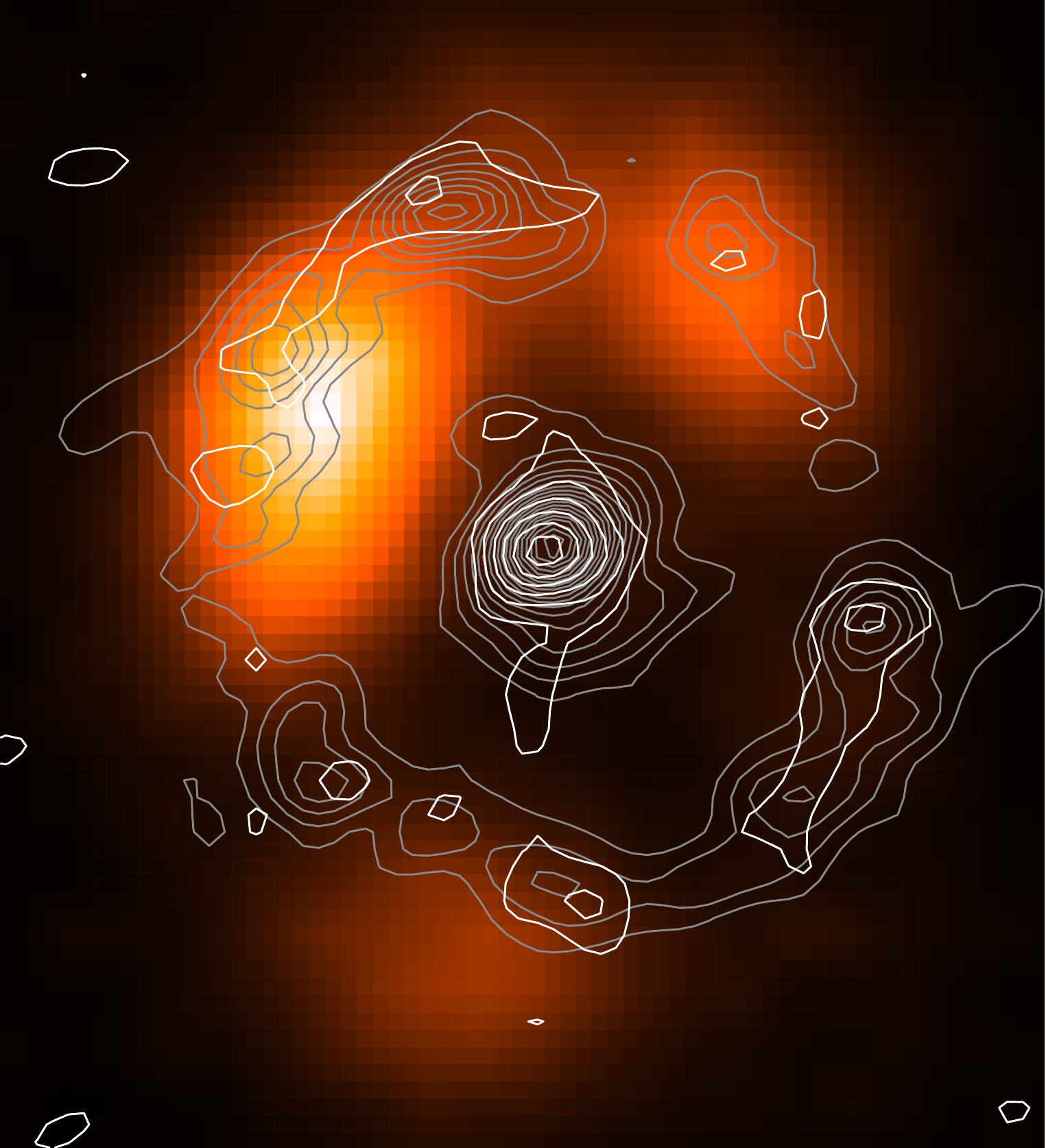}
\caption{
Comparison between the FUV emission \citep[in colour,][]{GildePaz2007}, and the HCN (grey contours) and CCH (white contours).
\label{fig.FUVHCNCCH}}
\end{figure}

Fig.~\ref{fig.FUVHCNCCH} shows a comparison between the FUV emission from unobscured massive star formation in the SF ring of NGC~1097 \citep{GildePaz2007}
with our observed HCN and CCH emission.
The north-eastern region, where the brighter FUV emission is observed, does not correspond to the higher relative abundance of CCH.
There (positions B and C), we find a large fraction of dense well UV-shielded gas, therefore decreasing the CCH relative abundance with respect to HCN, even though its 
absolute column density is as high as at other positions of the ring. Additionally the bright FUV from stars already formed would photodissociate CCH.
On the other hand, the position I, also highly illuminated in the north-east, has significantly less dense molecular gas, but one of the highest CCH relative abundances
showing that both evolved (strong observed UV) and ongoing stage star forming (enhanced CCH) regions co-exist in this region.
That is similar to what is observed in position G towards the south. There, though the observed FUV is not so intense, likely due to extinction,
it is enough to photo-enhance the abundance of CCH in a more diffuse medium, resulting in the highest CCH relative abundance measured.
This scenario is similar to that proposed to explain the high CCH abundance in the isolated galaxy CIG~638 \citep{Martin2014}.
Finally, towards the south-west (positions E, and F), we find a region where marginal FUV is observed, again possibly due to obscuration in this gas-rich
environment.
If CCH is indeed mostly enhanced in PDRs associated with massive star forming hot cores, this would be indicative of obscured star formation in this region, which is consistent
with the $860~\mu$m emission observed in these positions, also claimed to be tracing dust massive star forming regions \citep{Izumi2013}.

Though CCH can be a potential tracer of obscured star formation, its abundance variations can also be probing changes in the diffuseness of the ISM and therefore its permeability
to the UV irradiation from massive stars.
It is important to note that the critical density of CCH is about an order of magnitude below than that of species such as HCN so excitation effects might again play an
important role in the relative abundance derived, particularly towards the densest cores.

\subsection{The densest gas: SiO, HNCO and HC$_3$N}
\label{Sec.Densest}
These three species are the ones showing the clearest contrast between the central region and the SF ring with a significant enhancement
towards the nuclear region.
Towards NGC~1068, an enhancement of SiO and HC$_3$N in the surroundings of the AGN was already been recently reported \citep{Garc'ia-Burillo2010,Takano2014}.

SiO is only detected in one position outside the nuclear region, position E, and there it shows a relative abundance $25\%$ lower in this position.
Similarly, HNCO shows an underabundance of $30-40\%$ in the SF ring with respect to the central position, but for position C, where HNCO shows an enhancement of 20$\%$.
Moreover the limits towards the undetected positions indicate abundances $40-80\%$ below than that in the AGN vicinity.
Both species are considered as tracers of shocked dense material where shocks release both SiO and HNCO from the dust grain surfaces
\citep{Martin-Pintado1997,Huettemeister1998,Zinchenko2000,Mart'in2008}.
The emission of both species is observed to be well correlated not only in Galactic sources, but it is also observed in nearby bright galaxies
\citep{Usero2006,Meier2005}.
Surprisingly, in position E, where SiO has been detected in the SF ring, the lowest limit to the abundance of HNCO is found. This is likely
linked to the different evolutionary stage of the embedded star forming cores where UV radiation or high-velocity shocks are destroying
HNCO \citep{Mart'in2008}.

With a critical density of $\sim10^5~\rm cm^{-3}$ \citep{Wernli2007}, HC$_3$N is tracing the densest regions within the GMC complexes in NGC~1097.
Where detected, the measured relative abundances in the SF ring are $25-50\%$ of that in the central position.
The upper limits in the other positions are still around $40-50\%$ below.
In the case of HC$_3$N an increase in the temperature of the gas towards the central region would enhance this difference.
Our calculated ratio of HC$_3$N/HCN=1.2 for $T_{\rm ex}=8~K $ (see Section~\ref{subsec.CD}) would increase by a factor of $5-6$ for excitation
temperatures of $\sim 20$~K.

\section{Molecular variations within the central 300 pc}
\label{Sect.AGN}

Fig.~\ref{fig.RatioHCNHCOp} clearly suggests observable variations in the HCN/HCO$^+$ line ratios within the CND even though the central region is only resolved
over a few synthesized beams. Molecular variations in the central few hundred parsec CND around the AGN in NGC~1068 have been reported both
at very high spatial resolution \citep[35~pc,][]{Garcia-Burillo2014,Viti2014} for HCN an HCO$^+$, and at a coarser spatial resolution 
of $300''\times170''$ for a wider sample of species \citep{Takano2014}.
Here we discuss the different morphologies of the observed species towards NGC~1097 and compare it to what is observed in NGC~1068.

\subsection{The HCN/HCO$^+$ gradient}
\label{Sect.AGNHCNHCOp}
\begin{figure}
\centering
\includegraphics[width=\linewidth]{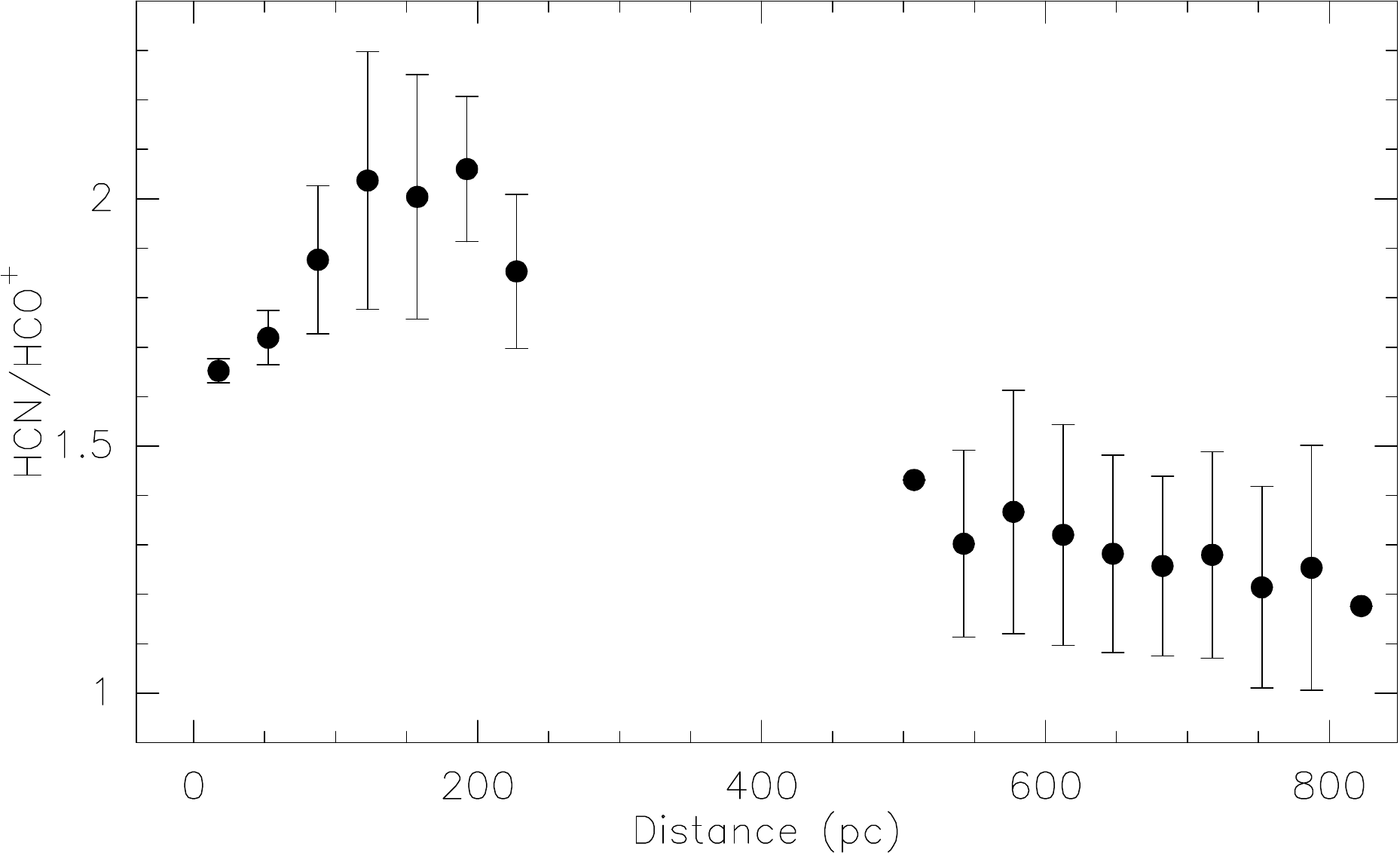}
\caption{
Integrated intensity HCN/HCO$^+$ ratio as a function of the galactocentric distance. Ratio has been averaged in bins of $0.5''$ (35~pc).
Error bars represent the dispersion of the values on each bin.
\label{fig.RatioHCNHCOpDist}}
\end{figure}

Within the central 300~pc we observe that the HCN/HCO$^+$ emission does not peak towards the AGN position. On the contrary, a positive gradient is observed as we get further
from the central position (Fig.~\ref{fig.RatioHCNHCOpDist}), reaching a relatively constant maximum value of $\sim2$. The ratio of $1.65$ at the central position
is still significantly higher than the average $\sim1.3$ measured in the SF ring (Sect.~\ref{HCN_HCO_SFAGN}).
The relatively low HCN/HCO$^+$ ratio in the close vicinity of an AGN compared to its surrounding CND has also been reported towards NGC~1068 \citep{Garcia-Burillo2014}.
Their 35~pc spatial resolution result in an even more dramatic gradient from a similar value towards the center of $\sim1.5$ up to a ratio of $\sim3$ in some CND regions.
As discussed above when comparing with the starburst ring, this gradient could be revealing the direct effect of the AGN imprinted in the molecular abundances on its immediate surroundings.
However, it has been claimed that the variation of this ratio could be the result of a density gradient, or due to differences
in the opacity of these species \citep[][Meier et al. submitted]{Meier2012}.

Though HCN and HCO$^+$ may be moderately optically thick (see Sect.~\ref{subsec.CD}), differential opacities are likely not to play a major role in the ratios observed where, 
for a similar column density, HCO$^+$ has an opacity almost twice that of HCN due to the higher radiative efficiency of HCO$^+$ as noted by \citet{Meier2012}. 
Given that both species show a similar excitation temperature \citep{Izumi2013}, 
a differential opacity effect might result in an increased HCN/HCO$^+$ ratio, where HCO$^+$ would reach saturation faster than HCN, if both species were similarly
enhanced towards the 
central position, contrary to what is observed.
On the other hand, the difference between their critical densities might suggest that HCO$^+$ is mostly concentrated towards the nuclear region where the densities are moderate 
$\sim 10^{4.5}~\rm cm^{-3}$, with a higher density region ranging $100-200$~pc around the AGN.
This scenario is equivalent to that discussed towards local starbursts \citep[][Meier et al. submitted]{Meier2012}, though the actual density structure and
morphology is likely to differ in both galaxy types.

The last scenario is a real relative abundance radial gradient from the AGN. However it could be either due to an enhancement of HCN production outwards from the nucleus or an inwards
overproduction of HCO$^+$.
The overproduction of HCO$^+$ due to enhanced star formation is discussed in Sect.~\ref{HCN_HCO_SFAGN} and ~\ref{Sect.SF}. 
Massive star formation could be taking place in the central 100~pc, or cosmic ray acceleration in the AGN would enhance HCO$^+$ but we have no evidence for any of these effects.
Thus, we rather claim that this gradient may be understood as a result of the enhancement of HCN in the material surrounding the circumnuclear disk (see Sect.~\ref{Sect.CNDHNCO}).
Based on theoretical \citep{PineaudesForets1990} and Galactic observations \citep{Tafalla2010}, the enhancement of HCN in the outflow of Mrk~231 has been claimed to be the result of shock chemistry
\citep{Aalto2012} where HCN is efficiently formed via the reaction CN + H$_2$.
We note that the HCN/HCO$^+$ enhancement is observed outside the CND defined by HNCO (Sect.~\ref{Sect.CNDHNCO}) and that the largest enhancement is observed in the extended HCN emission component emission.
Though HCO$^+$ shows extended emission south of the central region, HCN appears to show an even wider component extending in the north-east and south-west directions (Fig.~\ref{fig.allmolecules}).
However, we do not have other strong evidence of outflowing material in our images. On the contrary, the kinematic study of HCN by \citet{Fathi2013} shows evidences of inflowing
gas.

\subsection{The CND traced by HNCO}
\label{Sect.CNDHNCO}
Even though HNCO may be one of the most fragile species in our sample, it shows an enhancement towards the central region (Sect.~\ref{Sec.Densest}).
This species is the only one clearly delineating a rotating circumnuclear disk around the AGN position as shown in Fig.~\ref{fig.allmolecules}.
From the extreme emission peaks in the channel maps and taking into account the beam smearing, we measure a disk structure of $180\pm40$~pc length and P.A.$=120\pm10^\circ$,
covering a velocity range of 200~\kms~ around the systemic velocity. 
The measured velocity gradient of $1.1\pm0.2\rm~ km~s^{-1}~pc^{-1}$ is consistent with that measured towards the nuclear region by \citet{Hsieh2008}.
The orientation of the disk is marginally smaller than the kinematic P.A. of $133^\circ$, and $146^\circ$ derived from CO $2-1$ \citep{Hsieh2011} and HCN $4-3$, respectively \citep{Fathi2013}.
This might indicate a real variation between the overall rotation of the gas envelope traced by HCN and the inner disk traced by HNCO.
The possibility of HNCO tracing a bipolar outflow rather than a disk structure is unsupported by its kinematics being closely following those of 
CO and HCN at larger scales.
As observed in other species, the south-east region of the disk is about a factor of two brighter than
the north-west one. However the slightly narrower line width at this position ($\sim 80$~\kms) makes this difference not to be apparent in the integrated maps.
This is an indication of a significant asymmetrical distribution of the molecular gas in the CND \citep[also seen in CO $2-1$,][]{Hsieh2008}, similar to what is observed at higher resolution in NGC~1068 \citep{Garcia-Burillo2014}.
Also in NGC~1068, though at slightly lower resolution, the emission of HNCO is observed distributed in two lobes at both sides of the AGN \citep{Takano2014}. Such double peak distribution is in agreement with the CND being
observed from a more inclined direction consistent with its Seyfert 2 nucleus.

\begin{figure}
\centering
\includegraphics[width=\linewidth]{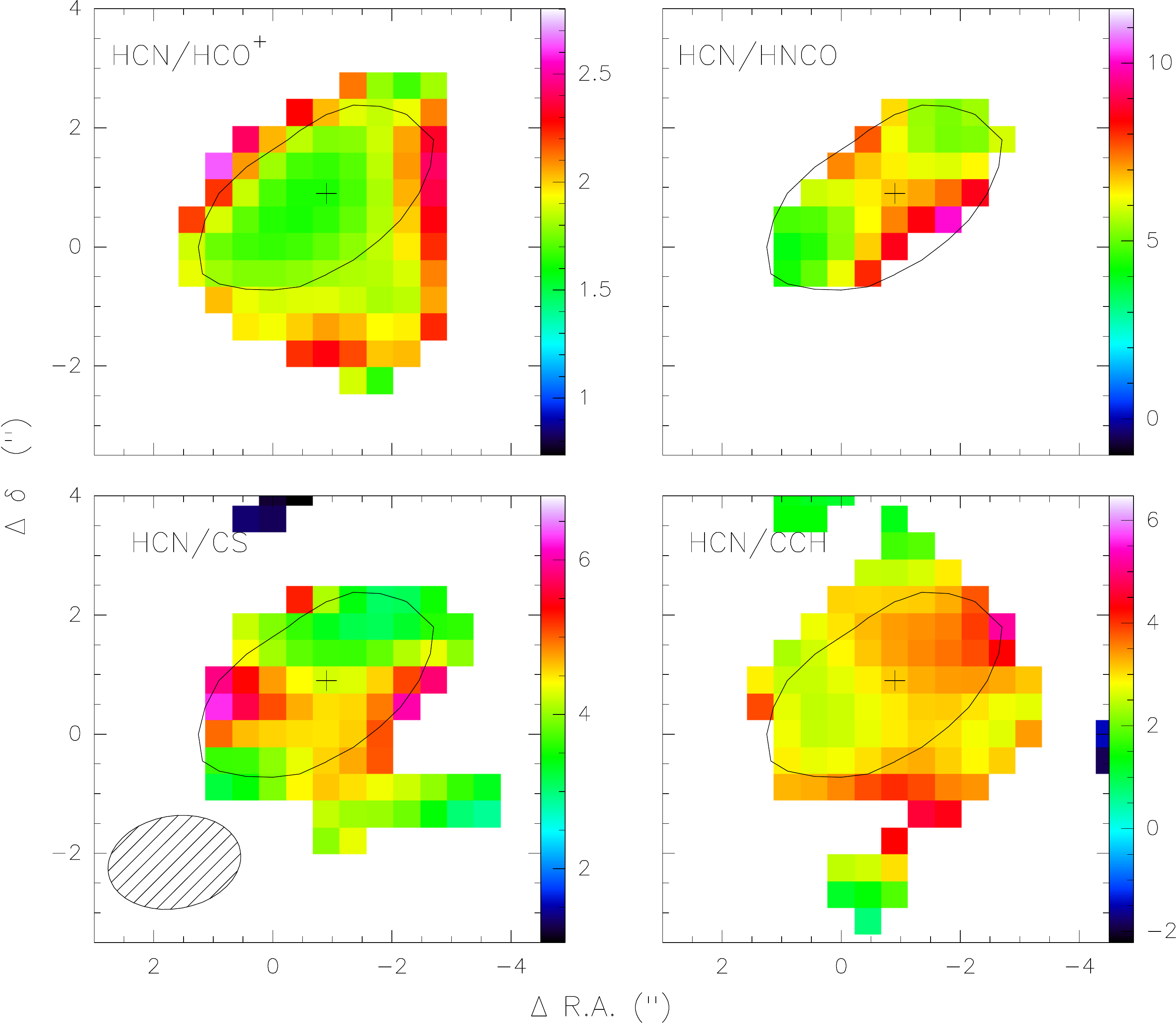}
\caption{
Integrated intensity line ratios of HCN over HCO$^+$, HNCO, CS and CCH in the central $\sim500$~pc of NGC~1097.
Position of peak of molecular emission is marked with a cross. The $3\sigma$ contour of HNCO emission, delineating the circumnuclear disk around the AGN, is overlaid as a reference.
Synthesized beam size is represented in the bottom left box.
\label{fig.molRatios}}
\end{figure}

HNCO forms in the very inner regions of molecular cloud, shielded from external UV irradiation by large visual extinction as shown by chemical models \citep[$A_v>4$~mag,][]{Mart'in2009b}.
This species is exclusively distributed along the circumnuclear rotating structure.
Within the central 300~pc, HNCO is absent in the regions outside the rotation plane of the CND, unlike the brightest species HCN, HCO$^+$, CCH and CS, which clearly show a wider extent.
In this region, low velocity shocks in the ISM due to cloud to cloud collision similar to what is observed in our Galactic center as well as in external galaxies 
\citep{Jones2012,Meier2005} will provide the environmental conditions for its ejection into gas phase while being dense enough to protect it from dissociation.
Though sensitivity could be claimed for such non detection, the HCN/HNCO ratio map in Fig.~\ref{fig.molRatios} shows already HNCO being exclusively concentrated in the disk, with the lower ratio
towards the two CND lobes, and a clear bipolar decrease in the perpendicular direction to the plane.
This fact points out to an increase of the radiation affecting the gas outside the CND. Given that star formation is not prominent in this region, UV generated in
high-velocity shocks might play a significant role in the destruction of HNCO, which may also explain the enhancement of HCN with increasing distance from the nucleus
(Sect.~\ref{Sect.AGNHCNHCOp}).
Additionally, an ionization cone due to a putative jet or direct exposure to the X-ray emission from the AGN,
could be claimed together to explain the absence of HNCO within the inner region ($<100$~pc) enclosed
by the CND and the biconical shape of the HCN/HNCO ratio map.

\subsection{HC$_3$N and SiO in the close vicinity to the AGN}
Both HC$_3$N and SiO are observed to be concentrated within an unresolved region (i.e. $\lesssim150$~pc) around the AGN. 
Such a high concentration of these molecules in the surroundings of an AGN has also been reported in NGC~1068 \citep{Garc'ia-Burillo2010,Takano2014}.

Though HC$_3$N shows some emission in the south-eastern position of the CND, the brightest in the CND, it is mostly concentrated in the central beam. 
\citet{Takano2014} reported that a $\sim  70\%$ of the HC$_3$N emission stems from the CND in NGC~1068.
Similar to what is observed in NGC~1097, HC$_3$N is more concentrated around the AGN than molecules tracing the CND such as HNCO.
Our observations together with those in NGC~1068 provide the evidence of the existence of a large amount of high density gas ($\sim 10^{6}~\rm cm^{-3}$) shielded from the X-ray radiation from the AGN,
as described by theoretical models from \citet{Harada2013}. In those models, a high temperature region of the disk moderately affected by X-rays can explain an enhancement
of the cyanopolyynes. In this scenario, HC$_3$N might be tracing the inner CND directly enclosed by HNCO.
Within scales almost two orders of magnitude smaller, the few central parsecs around the Galactic black hole show HC$_3$N (and to a large extent SiO) to be tracing the very densest
clumps in the circumnuclear disk material \citep{Mart'in2012}.

HC$_3$N has been observed to be overluminous in the deeply buried IR luminous galaxy NGC~4418 \citep{Aalto2007a} which might be a signature of either buried star formation \citep{Bayet2008a} or
contribution from gas in the vicinity of the AGN.
HC$_3$N enhancement has been confirmed in other obscured LIRGs \citep{Costagliola2011}.
The observed HC$_3$N/HCO$^+$ integrated intensity ratio toward NGC~4418 is $\sim 0.4-0.6$ \citep{Aalto2007a,Costagliola2011}. This is using the $J=10-9$ transition of HC$_3$N, and therefore the ratio would be higher
if using the $J=11-10$ in our work.
The ratio we derive towards the nucleus of NGC~1097 is however $\sim0.17$, and therefore significantly lower than the value in NGC~4418.
Similarly, for those LIRGs where HC$_3$N was detected in the sample from \citet{Costagliola2011}, a ratio ranging HC$_3$N/HCN$\sim0.4-0.2$ is found (again using the fainter $J=10-9$ HC$_3$N
transition).
Towards NGC~1097 we observe a lower ratio of $\sim0.1$.
Though we know that both species are not coexistent in NGC~1097, and thus a higher ratio of HC$_3$N might be measured towards the central position, a similar non-coexistence applies in the 
obscured LIRGs observed with single-dish resolution.
Therefore, based on our observations, an enhancement of HC$_3$N in the CND around a putative AGN cannot be the only contribution to the observed HC$_3$N enhancement in obscured LIRGs.
This gives a further support to the idea of buried young star formation in this objects as a key ingredient driving their luminosities \citep{Aalto2007a,Costagliola2011}.
We note, however, that the line ratios HC$_3$N/HCO$^+\sim0.16$ and HC$_3$N/HCN$\sim0.1$  measured towards the close IR bright quasar, Mrk~231, are in quite good agreement to those measured
in the center of NGC~1097, even though the strong star formation in Mrk~231 \citep[][and references therein]{Aalto2012}.

On the other hand, in the CND region SiO is exclusively detected towards the central beam. Though the signal to noise of our data is not high, the SiO/H$^{13}$CO$^+>1$ line
ratio is consistent with the observations towards NGC~1068 \citep{Usero2004}. 
SiO is a tracer of shocks both in the Galaxy \citep{Martin-Pintado1992,Martin-Pintado1997,Huettemeister1998} and in external galaxies \citep{Garc'ia-Burillo2000,Usero2006}.
Towards NGC~1068, X-ray and cosmic ray chemistry has been claimed to explain the SiO abundance in its CND \citep{Garc'ia-Burillo2010, Aladro2013} based on the correlation between SiO emission and the Fe 6.4 keV line
in the Galactic center \citep{Mart'in-Pintado2000,Amo-Baladr'on2009}.
However, the high abundance of the easily destroyed HC$_3$N would not resist photodissociation in such environment. 
Still, the emission could be enhanced by the X-ray radiation in a region different to that where HC$_3$N is formed \citep{Harada2013}, which might explain the enhanced abundance
of both species.
The signal-to-noise ratio of our SiO detection prevents us from reaching further conclusions, but to confirm the compact SiO emission in the surroundings of AGN, similar
to what is observed in NGC~1068 \citep{Garc'ia-Burillo2010}.

\subsection{CS and CCH}
As shown in Fig.~\ref{fig.molRatios}, the behavior of these two species in the central few hundred parsecs appear to anticorrelate.
Though the resolution is limited, based on previous studies and theoretical models we suggest that such differentiation is likely to be due to real abundance differences caused by
environmental conditions.

CS is observed to be evenly distributed across the CND, with a relative abundance drop of almost a factor of two in the south-eastern CND region (observed peak of HCN/CS ratio in Fig.~\ref{fig.molRatios}),
contrary to what is observed in CCH.
Though its abundance varies, CS seems to be abundant in the CND and not prominent in the region outside that plane.

Our observations show the CCH to be more abundant (lower HCN/CCH ratio in Fig.~\ref{fig.molRatios}) towards the south-east CND region where the peak line emission is found for most species,
with a significant but not too prominent ($\sim 60\%$) relative abundance decrease towards the north-western part of the CND.
In addition its abundance is also significant outside of the CND plane defined by HNCO, which might be linked to outflowing material similar to what is observed in Maffei~2 \citep{Meier2012}.
The differences observed might be due to temperature affecting its chemistry.
The models from \citet{Harada2013} (for the smaller XDR fraction to better represent this low-luminosity AGN) show CCH abundance to decrease as the temperature increase.

\section{Summary}

In this paper we have presented the 3~mm molecular observations carried out with ALMA towards NGC~1097.
The detection of eight different molecular species has been discussed under our current understanding of the formation and excitation of each of these species.
We have developed a molecular scenario of NGC~1097 first between the star forming ring and the nuclear region and secondly regarding the barely resolved morphology of
the emission of these molecules in the central $\sim300$~pc

The highest concentration of dense gas is located in the surroundings of the AGN \citep{Hsieh2012,Izumi2013}. 
The species HCN, HCO$^+$ and CS appear to be tracing the same molecular component. An increase of the abundance of HCO$^+$ and CS relative to HCN of $20-50\%$ in the star forming ring
with respect to the central region is observed.
The claimed large discrepancy in the HCN/CS ratio between SB and AGN galaxies \citep{Izumi2013} is likely due to a real underabundance of CS of a factor of $\sim2$ together with
excitation differences. CS is observed to be excited to higher temperatures \citep{Aladro2011}, which may be the evidence of a more prominent high density component
($10^6\rm cm^{-3}$) in SBs and may be less significant in AGNs. Additionally, CS could be underproduced in this high density regions in AGNs.

CCH shows overabundances of up to $70\%$ in the SF ring, and its abundance distribution differs significantly from the other species.
We claim it to be closely related to star formation and it might be a good tracer of obscured star formation in the ring.
The largest abundance contrast is shown by the densest gas tracers, namely SiO, HNCO, and HC$_3$N, which are clearly overabundant around the AGN.

Within the central $\sim300$~pc HNCO clearly delineates a circumnuclear disk of $180\pm40$~pc around the nucleus. The disk P.A.$=120\pm10^\circ$ is slightly less inclined
than the rotation plane of the overall molecular gas in the central region \citep[P.A.$=133-146^\circ$,][]{Hsieh2011,Fathi2013}.
This disk appear to enclose the emission of HC$_3$N and SiO which suggest an stratification of the disk as predicted by models \citep{Harada2013} with each species being formed/enhanced
at different regions within the disk.
In this scenario, HNCO would trace the core of the molecular clouds in the disk well protected from the radiation from the AGN. Inside we would find a dense region moderately exposed
to X-ray radiation where HC$_3$N would be enhanced. Finally the radiation resilient SiO might be tracing the innermost molecular region of the circumnuclear disk.

HCN/HCO$^+$ shows a positive gradient as we move away from the AGN, similar to what is observed towards NGC~1068, and therefore an enhancement of HCN via X-ray from the AGN is
discarded. Unless HCO$^+$ has some mechanism to be enhanced in the close vicinity of the AGN, we find plausible the HCN enhancement off the circumnuclear disk plane due to
shocked gas in outflowing material from the central engine.

The results presented in this paper towards the low luminosity AGN in NGC~1097 are in agreement with previous results on the luminous AGN on NGC~1068 \citep{Garcia-Burillo2014,Takano2014},
thus the difference in luminosity between the two sources does not appear to have a impact on its chemical structure around their AGNs.
These early science ALMA results clearly show the potential of chemical studies to understand the physical processes in the circumnuclear disks around AGNs.

\begin{acknowledgements}
This paper makes use of the following ALMA data: ADS/JAO.ALMA\#2011.0.00108.S. ALMA is a partnership of ESO (representing its member states), NSF (USA), and NINS (Japan),
together with NRC (Canada) and NSC and ASIAA (Taiwan), in cooperation with the Republic of Chile. The Joint ALMA Observatory is operated by ESO, AUI/NRAO, and
NAOJ. 
\end{acknowledgements}

\bibliographystyle{aa}	
\bibliography{1097ALMAB3.bib}	

\end{document}